\documentclass[twocolumn, showpics, preprintnumber, amsmath,amssymb]{revtex4}
\usepackage{graphicx}
\usepackage{dcolumn}
\usepackage{bm}
\begin{document}
\preprint{APS/123-QED}

\title{Extraordinary features of pair plasma and \\quasilinear theory of
Cherenkov-drift instability}

\author{David Shapakidze}
\altaffiliation[Also at ]{Abdus Salam International Centre for Theoretical Physics, Trieste}
\affiliation{International Center for Dense Magnetized Plasma, Institute of
Plasma Physics and Laser Microfusion, Hery 23, 01-497, Warsaw, Poland}

\author{George Machabeli}
\affiliation{Center for Plasma Astrophysics, Abastumani
Astrophysical Observatory, Al.Kazbegi ave. 2a, Tbilisi 380060,
Georgia.}

\author{George Melikidze}
\email{gogi@astro.ca.wsp.zgora.pl} \affiliation{Institute of
Astronomy, University of Zielona G\'ora, Lubuska 2, 65-265,
Zielona G\'ora, Poland}

\author{David Khechinashvili}
\affiliation{Institute of Astronomy, University of
Zielona G\'ora, Lubuska 2, 65-265, Zielona G\'ora, Poland}

\date{\today}

\begin{abstract}
We examine linear and quasi-liner stages of Cherenkov-drift
instability in the relativistic magnetized electron-positron
plasma. The external magnetic field lines are assumed to be
slightly curved. In this case the curvature drift of relativistic
beam particles plays decisive role in the development of the
instability. Quasi-linear relaxation of the relativistic beam
leads to diffusion of the resonant particles in the momenta space.
The expressions for diffusion coefficients of Cherenkov-drift
instability are obtained.
\end{abstract}


\pacs{52.35Qz, 41.75.Ht, 41.60.Bq, 95.30.Qd.}

\maketitle

\section{\label{sec:level1}Introduction}

Cherenkov-drift instability was suggested by Kazbegi, Machabeli \&
Melikidze \cite{kmm89,kmm91c,kmm92b,kmms96} as a possible
mechanism for generation of pulsar radio emission and later it was
approved in \cite{lmb99}. In those works the linear theory of
Cherenkov-drift instability was developed. It was shown, that in
the pulsar magnetosphere due to Cherenkov-drift instability the
orthogonally polarized plasma waves are excited. These waves can
escape from the magnetosphere and reach observer as a pulsar radio
emission. The necessary condition for the development of
Cherenkov-drift instability (as for an usual Cherenkov
instability) is a presence of a beam of particles in the
relativistic magnetized pair plasma (consisted of relativistic
electrons $e^{-}$ and positrons $e^{+}$).

Generally, Cherenkov type instabilities develop due to a resonant
interaction between waves and particles of a beam. The resonance
occurs, when the electric field vector ${\mathbf E}$ and the wave
vector ${\mathbf k}$ of generated waves have got components along
direction of the beam velocity ${\mathbf v}$ (${\mathbf
E}\cdot{\mathbf v}\neq 0$ and ${\mathbf k}\cdot{\mathbf v}\neq
0$). So, in the magnetized plasma the  transverse waves (${\mathbf
E}\perp{\mathbf B}\perp{\mathbf k}$) propagating along the
external magnetic field (~${\mathbf B}_0\parallel {\mathbf k}$~)
cannot be generated by an usual Cherenkov instability (because it
develops on the beam particles moving along the external straight
magnetic field lines: ${\mathbf v}\parallel{\mathbf B}_0$ and
${\mathbf E}\cdot{\mathbf v}= 0$ ).

Cherenkov-drift instability develops when the beam particles move
along slightly curved magnetic field (SCMF) lines and, hence,
drift across the plane where the curved lines lie. The drift
motion of beam particles provokes generation of purely transverse
as well as longitudinal-transverse waves.

Generally there are two most important effects caused by the the
particle relativistic motion along the SCMF line: curvature drift
and curvature radiation. The drift velocity is directed across the
plane of the SCMF lines and is given by the following expression:
\begin{equation}
u_d=\frac{\gamma v_{\parallel}^2}{\omega_B R_B}. \label{udr}
\end{equation}
Here $u_d$ denotes the drift velocity of electrons (positrons
drift with the same velocity but in the opposite direction);
$\omega_B= eB/mc$ is the cyclotron frequency of electrons; $R_{B}$
is the curvature radius of the magnetic field line; $\gamma$ is
Lorentz factor of a particle; $c$ is speed of light and
$v_{\parallel}$ is the component of ${\mathbf v}$ along the
magnetic field line. If $\gamma\gg 1$, the value of drift velocity
$u_d$ could be significant.

A single particle, moving along the curved field line, radiates so
called curvature radiation which can be easily described as a
synchrotron radiation in an effective magnetic field (see e.g.
\cite{z96}). In 1975 Blandford \cite{b75} investigated the
curvature radiation of plasma flowing along the SCMF lines. The
problem was studied in the limit of infinite magnetic field
$B_0\rightarrow\infty$ and it was shown that there is no radiation
at all: the waves, radiated by each particle, are absorbed by
another one. This result was confirmed later in papers
\cite{zs79,m80,cs88,lm92,kmm91c}, where spatially unlimited plasma
flow was considered. Then in the paper by Asseo, Pellat and Sol
\cite{aps83} the sharp boundary was assumed at the edge of the
flow propagating along the curved field line and the possibility
of the waves excitation was shown at this boundary. If plasma flow
has zero width the instability is reduced to that of
Goldreich-Keeley \cite{gk71}.

Development of Cherenkov type instability, taking into account the
curvature drift motion, was studied and growth rate was calculated
in \cite{kmm86,kmm89,kmm91c,kmms91,kmm92a,kmm92b,pmt92,pmt94} for
different particular cases. These results were confirmed later
after thorough investigation of the problem in \cite{lmb99,lbm99}.
The instability was called a Cherenkov-drift instability.

Presence of the curved magnetic field lines is the necessary
condition for both the Cherenkov-drift radiation in plasma and the
single particle curvature radiation in vacuum. However, the
Cherenkov-drift radiation still could not be interpreted as a
plasma curvature radiation, analogous to the single particle
curvature radiation: as it will be shown below, the
Cherenkov-drift radiation is not generated in the case of
$u_{d}\rightarrow 0$. On the other hand, it is evident that the
drift velocity equals to $u_{d}\approx 0$ if we assume that
$B_0\rightarrow\infty$. However a single particle radiates even
for the infinite intensity of magnetic field. Moreover, the single
particle radiates the vacuum wave, while the proper waves of the
medium (i.e. relativistic electron-positron plasma) are generated
in the case of Cherenkov-drift instability. This particular point
was not considered in the works by Blandford \cite{b75} and
Melrose \cite{m80}. Polarization of these waves strongly differs
from that of vacuum waves.

Brief examination of the linear theory of Cherenkov-drift
instability is discussed in section 2. In section 3 the
quasilinear equations for the Cherenkov-drift instability are
obtained. In section 4 coefficients describing diffusion of
particles in momentum space are evaluated. Alteration of plasma
distribution function is studied. The results are summarized in
section 5.

\section{The Linear Theory}

Properties of relativistic electron-positron plasma were carefully
investigated in series of works \cite{hr78, vkm85,ab86,lmmp86} and
are still the subject of concern \cite{gmg98,mg99}. The
electron-positron plasma of pulsar magnetosphere consists mainly
of two components: a bulk of plasma particles with high density
and low Lorentz factors ($\gamma_p \sim 3\div10$) and a beam of
the primary particles ejected from the stellar surface. Density of
the beam $n_b \sim n_{GJ}$, where $n_{GJ}$ is so called
Goldreich-Julian density. Density of secondary plasma $n_p \sim
\Sigma \cdot n_b$, where $\Sigma$ is the Sturrock 'multiplication
factor' ($\Sigma \sim 10^2 \div 10^7$) \cite{s71}.

\begin{figure*}
\includegraphics{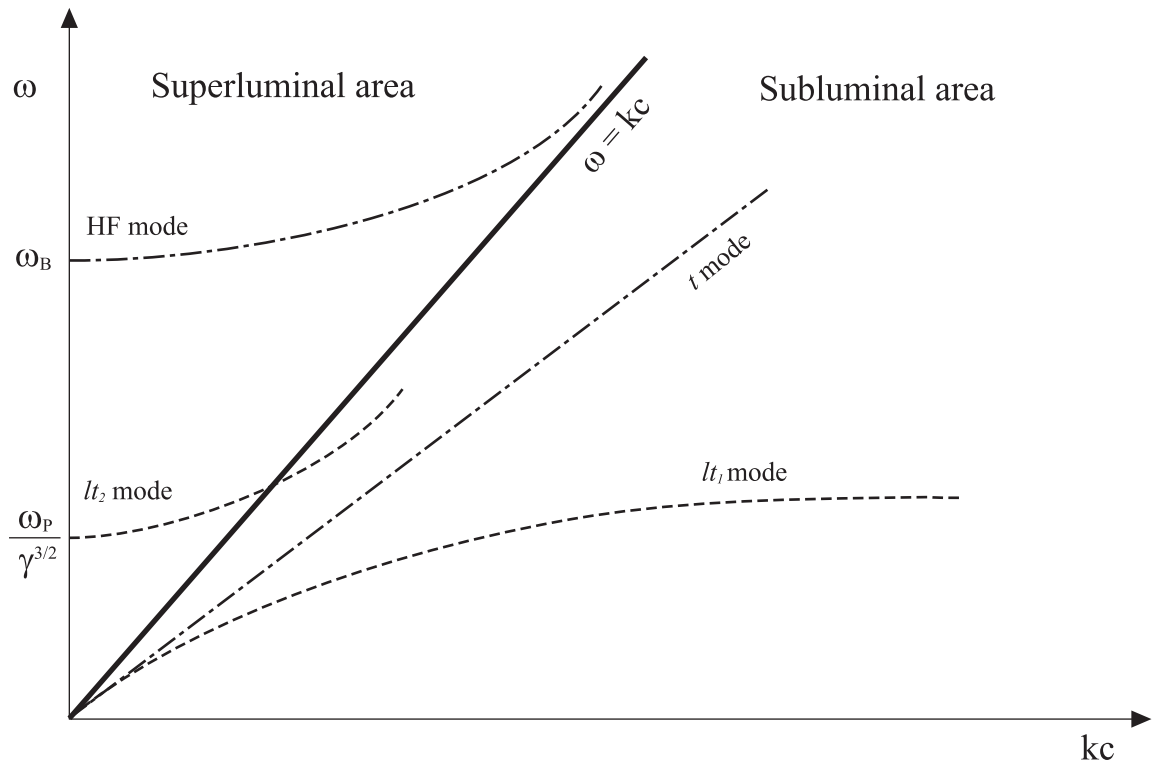}
\caption{\label{spectr} Dispersion curves of the waves in the
relativistic magnetized electron-positron plasma. The solid line
corresponds to $\omega=kc$. It divides the plane ($\omega,kc$)
into superluminal and subluminal areas. The high-frequency branch
of $t$ waves (in the superluminal area) is defined as HF mode.
Low-frequency and high-frequency modes of $lt$ waves are defined
as $lt_1$ and $lt_2$ modes respectively.}
\end{figure*}

The electron-positron plasma differs from the electron-ion plasma
by lack of gyrotropy. Consequently, spectra of waves propagating
in the $e^{-}e^{+}$ plasma is simpler than that in the
electron-ion plasma. It consists of only four types of waves which
correspond to four branches on the diagram $\omega(k)$, where
$\omega$ is the wave frequency (see Fig.\ref{spectr}).

One of them is high-frequency transverse electromagnetic wave
totally located in superluminal area. Its phase velocity
$v_{ph}=\omega/k$ exceeds speed of light $c$, hence, it is not of
interest in our discussion below.

The second branch represents the dispersion of purely transverse
linearly polarized electromagnetic wave. We call this dispersion
curve $t$ mode. It is totally located in subluminal area and
therefore could be generated by beam particles. Its electric field
vector ${\mathbf E}^t$ is perpendicular to the plane of wave
vector and external magnetic field (${\mathbf k},{\mathbf B}_0$).

The remaining two dispersion curves on the $ \omega(k)$ diagram
describes the longitudinal-transverse waves propagating in
relativistic $e^-e^+$ plasma. One of them is almost superluminal.
This wave is purely longitudinal if it propagates strictly along
the magnetic field line (${\mathbf k}\parallel {\mathbf
E}\parallel {\mathbf   B}_0$), and, in this case called Langmuir
wave associated with longitudinal oscillations of the charge
density. If an angle $\vartheta$ between ${\mathbf k}$ and
${\mathbf B}_0$ increases, the component of wave electric field
${\mathbf E}$ starts to grow across ${\mathbf k}$: Langmuir wave
transforms to the longitudinal-transverse wave denoted in
Fig.\ref{spectr} as $lt_2$ mode. If the angle $\vartheta$ is small
enough ($\vartheta \leq \vartheta_0 \sim
\omega_p^2/\omega_B^2\gamma_p^3$, where $\omega_p=\sqrt{4\pi
n_{p}e^2/m}$ is the Langmuir frequency), $lt_2$ mode is almost
longitudinal and crosses $\omega=kc$ line. In this case, $lt_2$
mode could be excited, if Cherenkov resonance condition
$\omega=k_{\parallel}v_{\parallel}$ is fulfilled. However, for the
resonant particles of primary beam, growth rate of the instability
is very small \cite{elm83}. The wave leaves from the interaction
area so quickly that no time is left for significant growth of the
wave amplitude. In the case of oblique propagation, $\vartheta
> \vartheta_0$, $lt_2$ mode is totally superluminal. Therefore,
$lt_2$ mode could not be generated at all by particles of beam.

Another longitudinal-transverse wave is denoted in
Fig.\ref{spectr} as $lt_1$ mode. This mode, like $t$-wave, is
located totally in subluminal area and can easily be generated by
plasma particles. Its electric field vector ${\mathbf E}^{lt}$ is
located in (${\mathbf k},{\mathbf B}_0$) plane. $lt_1$ mode is
vacuum wave, if it propagates along the magnetic field lines
(${\mathbf k}\parallel{\mathbf  B}_0$). Its dispersion curve
merges with $t$ mode (see Fig.\ref{spectr}) and can be arbitrarily
polarized. In the case of oblique propagation, electric field of
$lt_1$ wave has the component $E_{\parallel}^{lt}$ along the
external magnetic field, thereby involving plasma particles in
longitudinal oscillations.

Generation of $lt_1$-mode, propagating in perpendicular direction
to the plane of SCMF lines, is connected with the drift motion of
the particles. These waves are also known as \emph{drift waves}
\cite{kmm91c,kmms96,clmms97}.

It should be mentioned that, while describing the waves in
relativistic $e^{-}e^+$ plasma, some authors sometimes use
terminology which, in our opinion, appears to be misleading. For
example, since the work by Arons and  Barnard \cite{ab86}, the
superluminal longitudinal-transverse wave ($lt_2$ mode) was called
an ordinary ($O$) mode, the subluminal transverse wave ($t$ mode)
-- an extraordinary ($X$) mode and the subluminal
longitudinal-transverse wave ($lt_1$ mode) -- an Alf\'{v}en mode.
However 'ordinary' and 'extraordinary' are generally related to
the waves propagating across the external magnetic field in usual
electron-ion plasma \cite{kt73}. Moreover, $t$ wave (so called $X$
mode) is the purely transverse wave and its analogue does not
exist in the electron-ion plasma. In electron-positron plasma the
terminology \emph{ordinary} and \emph{extraordinary} are used for
the waves propagating almost along the magnetic field. As for the
Alf\'{v}en mode, in electron-ion plasma such a name is used for an
almost linearly polarized, transverse electromagnetic wave with
frequency $\omega \ll \omega_{B_i}$ (where $\omega_{B_i}$ is the
cyclotron frequency of ions). In the case of $k\rightarrow\infty$
and $\omega\rightarrow\omega_{B_i}$, this wave transforms into
right-hand polarized ion-cyclotron mode with frequency
$\omega\approx\omega_{B_i}$ and left-hand polarized
electron-cyclotron wave with frequencies
$\omega\approx\omega_{B_e}$. The latter case corresponds to the
fast magnetosonic wave and is called \emph{helicon} for
frequencies $\omega_{B_i}<\omega<\omega_{B_e}$. Therefore, in
relativistic $e^-e^+$ plasma, we prefer to call the dispersion
curves $t$, $lt_2$ and $lt_1$ modes respectively, hence avoiding a
possible confusion with dispersion curves in electron-ion plasma.

\begin{figure*}
\includegraphics{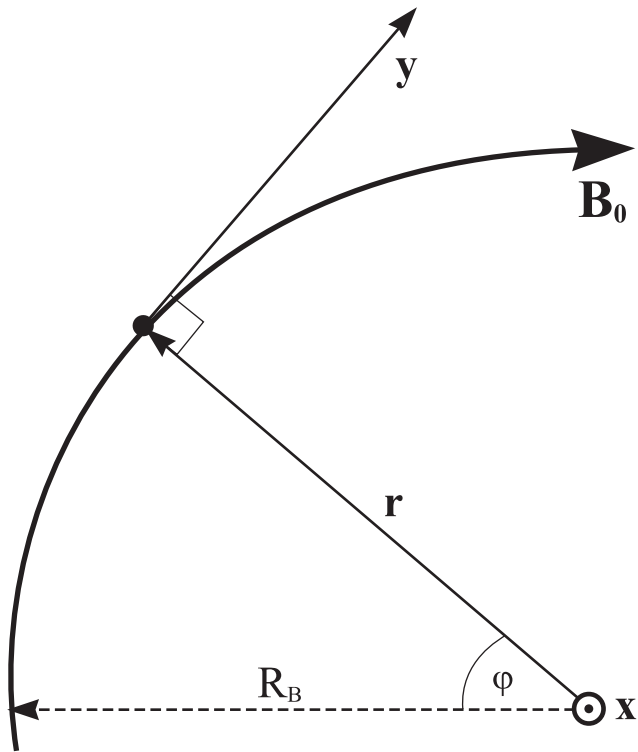}
\caption{\label{l2} The cylindrical frame
of reference ($x,\, r,\, \varphi$) and a local Cartesian frame of
reference ($x,\, r,\, y$). $x$ axis is directed up from the plane of the
figure. ${\mathbf B}_0$ is the external magnetic field; $R_B$ is
the radius of curvature of a magnetic field line.}
\end{figure*}

To obtain expression for dielectric permittivity tenor components
$\epsilon _{ij}(\omega, {\mathbf k})$ in the case of SCMF lines,
it is handy to consider the problem in cylindrical coordinates
$(x,\,r,\,\varphi)$ \cite{kmm91a} and direct $x$ axis
perpendicularly to the plane of the curved magnetic field lines
(see  Fig.\ref{l2}).
In the papers by Kazbegi, Machabeli and Melikidze
\cite{kmm91a,kmm91c,kmm92b}, it was shown that $t$ and $lt$ waves
could be generated by particles of the beam when the following
resonance condition is satisfied (for the resonant values of
parameters $\omega=\omega_0;\; v_{\varphi}=v_0\; \text{and}\;
u_d=u_0$):
\begin{equation}
\Delta\omega\equiv\omega-k_{\varphi}v_{\varphi}-k_x u_d = 0.  \end{equation}

Mechanism of wave generation is a modification of the well-known
beam-plasma instability. However, this instability differs
significantly from the usual beam-plasma instability
\cite{kmms96}.

The expressions for the
growth rates of the different waves are as follows:
\begin{eqnarray}
\Gamma_{\mathbf k}^t=\frac{\pi }{2}\frac{\omega _b^2}{\omega
_{\mathbf k}}\frac{\gamma _b}{\gamma _T^2}\frac{k_r^2}{k_{\perp
}^2}\label{Gammat}
\end{eqnarray}
for transverse ($t$) wave and
\begin{eqnarray}
\Gamma_{\mathbf k}^{lt}=\frac{\pi }{2}\frac{\omega _b^2}{\omega
_{\mathbf k}}\frac{\gamma_b}{\gamma_T^2}\frac{k_x^2}{k_{\perp
}^2} \label{Gammalt}
\end{eqnarray}
for longitudinal-transverse ($lt$) wave (here $\omega_b$ is
Langmuir frequency of resonant particles of the beam). These
expressions were obtained in papers \cite{kmm89,kmms91,kmm92b}. As
for growth rate of the drift wave, it was obtained in paper
\cite{kmm91c}:
\begin{equation}
\Gamma_d=\left(\frac{3}{2}\frac{n_b}{n_p}\right)^{1/2}
\left(\frac{\gamma_p^3}{\gamma_b}\right)^{1/2}k_x u_d.
\label{Gammadr}
\end{equation}

The reason for generation of $t$, $lt$ and drift waves is presence
of the beam in the relativistic pair plasma, although the waves
cannot be excited without drift motion of particles. Indeed,
expressions (\ref{Gammat}), (\ref{Gammalt}) and  (\ref{Gammadr})
are equal to zero if $u_d=0$. All those waves (see
Eqs.~\ref{Gammat}-\ref{Gammadr}) are purely or almost transverse
waves. For $t$ and $lt_1$ waves the electric field vector is
perpendicular to the external magnetic field (see
Eqs.~\ref{Gammat},\ref{Gammalt}). As for the waves generated by
the usual beam instability, both their electric field vector
${\mathbf E}$ and the wave vector ${\mathbf k}$ are directed along
the external magnetic field ${\mathbf B}_0$.

In the following section we study the quasilinear equations which
are significantly different from those of the usual beam-plasma
instability.

\section{Quasi-linear Equations}

Let us study the quasilinear theory of Cherenkov-drift
instability, using the collisionless kinetic equation in the
following form:,
\begin{equation}
\frac{\partial f_{\alpha}}{\partial t}+{\mathbf v}\frac{\partial
f_{\alpha}}{\partial {\mathbf r}}+\frac{\partial}{\partial
{\mathbf p}} \left[\frac{q_{\alpha}}{m_{\alpha}c}\left( {\mathbf
E}+\frac{{\mathbf p}\times{\mathbf B}}{\gamma}\right)\cdot
f_{\alpha}\right]=0.\label{f}
\end{equation}
Here $f_{\alpha}\equiv f_{\alpha}({\mathbf r},{\mathbf p},t)$ is
the distribution function of particles of sort $\alpha$, ${\mathbf
E}({\mathbf r},t)$ and ${\mathbf B}({\mathbf r},t)$ are the
electric and magnetic field vectors, respectively.

According to standard scheme, in order to obtain a system of
quasilinear equations, distribution function, as well as electric
and magnetic field vectors, have to be divided into the main and
oscillating parts: $f_{\alpha}({\mathbf r},{\mathbf
p},t)=f_{\alpha 0}({\mathbf p}, \mu t)+f_{\alpha 1}({\mathbf
r},{\mathbf p},t)$, ${\mathbf E}({\mathbf r},t)={\mathbf E}_1
({\mathbf r},t)$ and ${\mathbf B}({\mathbf r},t)={\mathbf B}_0
({\mathbf r},\mu t)+{\mathbf B}_1 ({\mathbf r},t)$. Then,
averaging equation (\ref{f}) over fast oscillations and assuming
$\langle f_{\alpha 1} \rangle =\langle {\mathbf B}_{1} \rangle
=\langle{\mathbf E}_{1} \rangle =0$, $\langle f_{\alpha 0} \rangle
=f_{\alpha 0 }$ and $f_{\alpha 0} \gg f_{\alpha 1}$, the following
equations can be obtained:
\begin{subequations}
\begin{equation}
\frac{\partial f_{\alpha 0}}{\partial \mu t}= -\left\langle
\frac{q_{\alpha}}{m_{\alpha}c}\frac{\partial }{\partial{\mathbf
p}}\left( {\mathbf E}_1+\frac{{\mathbf p}\times{\mathbf
B}_1}{\gamma}\right)\cdot
 f_{\alpha 1}\right\rangle \equiv QL,
\label{f0}
\end{equation}
\begin{eqnarray}
\frac{\partial f_{\alpha 1}}{\partial t}&+& \frac{c{\mathbf
p}}{\gamma}\frac{\partial f_{\alpha 1}}{\partial {\mathbf
r}}+\frac{q_{\alpha}}{m_{\alpha}c} \left(\frac{{\mathbf
p}\times{\mathbf B}_0}{\gamma}\right) \frac{\partial f_{\alpha
1}}{\partial {\mathbf p}}\nonumber\\ = &-&\frac{q_{\alpha}}{m_{\alpha}c}\left(
{\mathbf E}_1+\frac{{\mathbf p} \times {\mathbf
B}_1}{\gamma}\right) \frac{\partial f_{\alpha 0}}{\partial{\mathbf
p}}. \label{f1q}
\end{eqnarray}
\end{subequations}
Here $\langle...\rangle$ denotes averaging over fast oscillations.
Equation (\ref{f0}) describes back reaction of the generated waves
upon the non-perturbed distribution function $f_{\alpha 0}$. In
order to calculate the quasilinear term $QL$ it is enough to
substitute the solution of Eq.~(\ref{f1q}) into Eq.~(\ref{f0}).
The solution of Eq.~(\ref{f1q}) in the Fourier presentation,
\begin{equation}  f_{\alpha 1}({\mathbf r},{\mathbf
p},t)=\frac{1}{(2\pi)^3} \int  f_{\alpha {\mathbf k}}({\mathbf
p})\exp (i\mathbf{kr}-i\omega_{\mathbf k} t)\,d{\mathbf k},
\label{f1wk}
\end{equation}
writes:
\begin{widetext}
\begin{equation}
f_{\alpha {\mathbf k}}({\mathbf p})=-\frac{q_{\alpha
}}{m_{\alpha}c}\int_{-\infty }^{t}dt'\exp[i{\mathbf
k}\bm{\rho}-i\omega _{\mathbf k}\tau]\left[{\mathbf
E}_{\mathbf k}\left( 1-\frac{{\mathbf p}'}{\gamma}\frac{{\mathbf
k}c}{\omega_{\mathbf k}}\right)+\frac{({\mathbf E}_{\mathbf
k}\cdot{\mathbf p}')}{\gamma}\frac{{\mathbf k}c}{\omega_{\mathbf
k}}\right]\frac{\partial f_{\alpha 0}}{\partial {\mathbf p}'},
\label{f1kwint}
\end{equation}
\end{widetext}
where $\omega_{\mathbf k}\equiv\omega({\mathbf
k})+i\Gamma_{\mathbf k}$; $\bm{\rho}={\mathbf r}'-{\mathbf r}$ and
$\tau=t'-t$. Then, perturbed electric field vector ${\mathbf E}_1$
is substituted for perturbed magnetic field ${\mathbf B}_1$ using
Maxwell equation ${\mathbf B }_{\mathbf k}=(c/{\omega}_{\mathbf
k})({\mathbf k}\times {\mathbf E}_{\mathbf k})$ for the Fourier
transforms:
\begin{eqnarray}
{\mathbf E}_1({\mathbf r},t)=\frac{1}{(2\pi)^3}\int{\mathbf
E}_{\mathbf k}\exp (i\mathbf{kr}-i\omega_{\mathbf k}
t)\,d{\mathbf k};  \nonumber \\
{\mathbf B}_1({\mathbf r},t)=\frac{1}{(2\pi)^3}\int{\mathbf
B}_{\mathbf k}\exp (i\mathbf{kr}-i\omega_{\mathbf k} t)\,d{\mathbf
k}.\label{Ewk}
\end{eqnarray}

In order to find $f_{\alpha {\mathbf k}}({\mathbf p})$, we are
using standard method of integration along non-perturbed
trajectories. Following this method, in Eq.~(\ref{f1kwint}),
(${\mathbf r}',\,{\mathbf p}'$) are phase coordinates of the
particle (along the non-perturbed trajectory) at the instant of
time $t'$; they are calculated from the relativistic equations of
motion of a single particle:
\begin{subequations}
\begin{equation}
\frac{d{\mathbf r}'}{dt'}=\frac{c}{\gamma}{\mathbf p}',
\label{eqmot1}
\end{equation}
\begin{equation}
\qquad \frac{d{\mathbf
p}'}{dt'}=\frac{q_{\alpha}}{m_{\alpha}c}\left(\frac{{\mathbf
p}'\times {\mathbf B}_0({\mathbf
r}',t')}{\gamma}\right).\label{eqmot2}
\end{equation}
\end{subequations}
Using cylindrical coordinate system, the non-perturbed magnetic
field is modeled as ${\mathbf B}_0={\mathbf
B}_0(0,\,0,\,B_{\varphi})$. It is supposed that non-perturbed
external electric field is absent in the plasma, ${\mathbf
E}_0=0$. The drift velocity (see Eq.~\ref{udr}) is directed along
$x$ axis. The particles of opposite charge $q_{\alpha}$ drift in
opposite directions. Hereafter the positive direction of $x$ axis
is the direction of the positive charge drift.

In such a geometry, the relativistic equation of motion
(\ref{eqmot2}) is rewritten as the following set of equations:
\begin{eqnarray}
\frac{\partial p_{x}'}{\partial t'}-\frac{\omega_{B_\alpha}}{\gamma}p_{r}'=0, \nonumber \\
\frac{\partial p_{r}'}{\partial t'}-p_{\varphi}' \frac{d
{\varphi}'}{d t'}+\frac{\omega_{B_\alpha}}{\gamma}p_{x}'=0, \nonumber \\
\frac{\partial p_{\varphi}'}{\partial t'}+p_{r}'\frac{d
{\varphi}'}{d t'}=0. \label{cyleqmot}
\end{eqnarray}
Ignoring the terms proportional to the small parameter
$(r_{L}/R_{B})^2$ (where $r_{ L}\simeq c/\omega_{B_\alpha}$ is the
radius of Larmor circle) and assuming that $p_{\varphi}^2 \gg
(p_r^2+p_x^2)$, we can obtain the solutions of set of
Eqs.~(\ref{cyleqmot}) as was done in \cite{pmt94}:
\begin{eqnarray}
p_{x}' &=& p_{d_\alpha}+p_{r} \sin
(\tilde{\omega}_{B_{\alpha}}\cdot\tau) +(p_{x}-p_{d_\alpha})\cos
(\tilde{\omega}_{B_{\alpha}}\cdot\tau), \nonumber \\
p_{r}' &=& p_{ r} \cos
(\tilde{\omega}_{B_{\alpha}}\cdot\tau)-(p_{x}-p_{d_\alpha}
)\sin(\tilde{\omega}_{B_{\alpha}
}\cdot\tau), \nonumber \\
p_{\varphi}' &=& p_{\varphi}. \label{solcyleqmot}
\end{eqnarray}
Here $\tilde{\omega}_{B_{\alpha }} = \omega_{B_{\alpha}}/\gamma$;
the components of dimensionless momentum ($p_{x}$, $p_{r}$,
$p_{\varphi}$) are the values of ($p_{x}'$, $p_{r}'$,
$p_{\varphi}'$) at the instant of time $t'=t$, and
$p_{d_\alpha}=(u_{d_\alpha}/c)\gamma$. Here we have the following
integrals of motion: $\gamma ,\; p_{x}'-\omega_{B_\alpha} r'/c,\;
r' p_{\varphi}'$. Let us notice that the particle distribution
function $f_{\alpha 0}$ should only depend on the integrals of
motion.

It is evident that the solutions of equation of motion
(\ref{solcyleqmot}) differ from those in the homogeneous magnetic
field only by the drift term $p_{d_\alpha}$. Using cylindrical
coordinates in momentum space
($p_{\parallel},\;p_{\perp},\;\theta$) (subscripts `$\parallel$'
and `$\perp$' denote parallel and perpendicular directions to the
magnetic field ${\mathbf B}_0$, respectively), we can write
\begin{eqnarray}
p_{\perp}\cos\theta&=&p_{x}-p_{d_\alpha}, \nonumber\\
p_{\perp}\sin \theta &=&p_{r}. \nonumber
\end{eqnarray}
Therefore Eqs.~(\ref{solcyleqmot}) are reduced to the following
form:
\begin{eqnarray}
p^{\prime}_{x} &=& p_{d_\alpha}+p_{\perp}\cos(\theta-\tilde{\omega}_{B_\alpha}\cdot\tau), \nonumber \\
p^{\prime}_{r} &=&p_{\perp}\sin(\theta-\tilde{\omega}_{B_{\alpha}} \cdot\tau), \nonumber \\
p_{\varphi}^{\prime} &=& p_{\parallel}. \label{solloceqmot}
\end{eqnarray}
Substituting Eqs.~(\ref{solloceqmot}) into Eq.~ (\ref{eqmot1}) we
obtain the following expressions for $\bm{\rho} = {\mathbf
r'}-{\mathbf r}$:
\begin{eqnarray}
\rho_x &=& \frac{cp_{d_\alpha}}{\gamma}\tau-
\frac{c}{\omega_{B_\alpha
}}p_{\perp}[\sin(\theta-\tilde{\omega}_{B_\alpha}\cdot\tau)-\sin\theta], \nonumber \\
\rho_r &=& \frac{c}{\omega_{B_\alpha
}}p_{\perp}[\cos(\theta-\tilde{\omega}_{B_\alpha}\cdot\tau)-\cos\theta],
\nonumber \\
\rho_{\varphi} &=& \frac{cp_{\parallel}}{\gamma}\tau,
\label{ro}
\end{eqnarray}
Substituting expressions (\ref{solloceqmot}) and (\ref{ro}) for
the components of $\bm{\rho}$ and ${\mathbf p }'$ in
Eq.~(\ref{f1kwint}), we finally obtain Fourier transform of
oscillating distribution function $f_{\alpha {\mathbf k}}({\mathbf
p})$:
\begin{widetext}
\begin{eqnarray}
f_{\alpha {\mathbf k}}({\mathbf p})=&-&
\left(\frac{q_{\alpha}}{m_{\alpha}c}\right)\frac{i\exp \left[
ib\sin (\theta-\varsigma) \right]}{\Delta\omega_{\alpha {\mathbf k}}}\nonumber
\\ &\times&\left[\,\frac{\partial f_{\alpha 0}}{\partial p_{\parallel }}
\left(E_{\parallel}({\mathbf k})+
\frac{k_{\parallel}c}{\omega_{\mathbf
k}}\frac{p_{d_\alpha}}{\gamma}E_{x}({\mathbf k})\right)
+ \frac{\partial f_{\alpha 0}}{\partial
p_{\bot }}\left(E_{\parallel}({\mathbf k})\frac{2p_{d_\alpha}\cos
\theta}{p_{\parallel}}+\frac{k_{\parallel}c}{\omega_{\mathbf k
}}\frac{p_{d_\alpha}}{\gamma}E_{x}({\mathbf
k})\frac{2p_{d_\alpha}\cos
\theta-p_{\bot}}{p_{\parallel}}\right)\right], \label{f1final}
\end{eqnarray}
\end{widetext}
where $\Delta\omega_{\alpha {\mathbf k}}\equiv \omega_{\mathbf
k}-k_{\parallel}v_{\parallel}-k_x u_{d_\alpha}$. Calculating
expression (\ref{f1final}) from Eq.~(\ref{f1kwint}), we used the
following presentation of the exponential function:
\begin{equation}
\exp[i{\mathbf k}{\bm \rho}-i\omega _{\mathbf
k}\tau]=\exp[ib\sin(\theta-\varsigma)]
\sum^{\infty}_{n=-\infty}{\cal E}_n\,J_n(b). \label{exp}
\end{equation}
Here ${\cal E}_n\equiv\exp[in(\varsigma-\theta)-
i\tau (\Delta\omega_{\alpha {\mathbf k}}-n\tilde\omega_{B_\alpha})]$;
$J_{n}(b)\; (n=0;\,\pm 1;\,\pm 2...)$ is the Bessel function  of integer order
\cite{as65};
\begin{equation}
b=\frac{k_\bot c}{\omega_{B_\alpha}}p_{\bot}
\label{b}
\end{equation}
and $\varsigma$ is defined as follows:
\begin{eqnarray}
k_{ x}&=&k_{\bot} \cos \varsigma,\nonumber \\
k_{r}&=&k_{\bot} \sin \varsigma. \label{dzeta}
\end{eqnarray}
In derivation of Eq.~(\ref{f1final}), we have assumed that $b\ll
1$ and kept only the terms with $n=0$ in Eq.~(\ref{exp}). This
approximation leaves only the terms describing contribution of
Cherenkov-drift resonance. Let us mention that $(\partial
f_{\alpha 0}/\partial \theta) = 0$ since the distribution function
possesses an axial symmetry. Therefore, corresponding terms do not
contribute into Eq.~(\ref{f1final}).

\section{Quasi-linear diffusion}

The next step is to study alteration of slowly varying part of
distribution function $f_{\alpha 0}({\mathbf p},\mu t)$ due to
development of Cherenkov-drift instability. Generally, alteration
is described by diffusion coefficients involved in the quasilinear
term. The coefficients show the rate of particle diffusion in
momentum space along, as well as across, the magnetic field.
Substituting Eq.~(\ref{f1final}) into Eq.~(\ref{f0}) and using
Maxwell equation for Fourier transforms, we obtain $QL$ term in
the following form:
\begin{eqnarray}
&&QL \equiv -\frac{q_{\alpha}}{m_{\alpha}c}\left\langle
\frac{\partial }{\partial{\mathbf  p}}\left( {\mathbf
E}_1+\frac{{\mathbf p}\times{\mathbf  B}_1}{\gamma}\right)
f_{\alpha 1} \right\rangle
\nonumber \\
&&=\frac{\partial}{\partial
p_{\parallel}}\left(D_{\parallel\parallel}\frac{\partial
f_{\alpha 0}}{\partial p_{\parallel }}\right)+
\frac{\partial}{\partial p_{\parallel}}\left( p_{\perp
}D_{\parallel\perp}\frac{\partial f_{\alpha 0}}{\partial p_{\perp
}}\right) \nonumber \\
&&+ \frac{1}{p_{\perp}}\frac{\partial }{\partial p_{\perp
}}\left( p_{\perp }^{2}D_{\perp
\parallel}\frac{\partial f_{\alpha 0}}{\partial p_{\parallel }}
\right)
+ \frac{1}{p_{\perp }}\frac{\partial }{\partial p_{\perp
}}\left( p_{\perp }D_{\perp \perp }\frac{\partial f_{\alpha
0}}{\partial p_{\perp }}\right)\,. \nonumber \\ \label{QL}
\end{eqnarray}
Here $D_{\parallel \parallel}$, $D_{\parallel \perp}$, $D_{\perp
\parallel}$ and $D_{\perp \perp }$ are the diffusion coefficients.
Below we calculate particular expressions for diffusion
coefficients corresponding to $t$ and $lt$ waves in the case of
Cherenkov-drift instability. Let us notice that Eq.~(\ref{QL}) has
rather general meaning and can as well be used for  other type of
instabilities. However, diffusion coefficients will differ for
different instabilities.

It is convenient to consider quasilinear development of $t$ and
$lt$ waves separately. Using simple geometrical assumptions, the
following relations between components of electric field and wave
vector can be written as
\begin{equation}
E^{t}_{\parallel}=0,\qquad \qquad E^{t}_{x}k_{x}=-E^{t}_{r}k_{r}
\label{Et,k}
\end{equation}
for purely electromagnetic (${\mathbf E} \perp {\mathbf k}$) $t$
waves and
\begin{equation}
E^{ lt}_{ x}k_{ r}=E^{ lt}_{r}k_{ x},\qquad \qquad E^{
lt}_{\parallel }k_{\parallel }=-E^{lt}_{ x}k_{x} \label{Elt,k}
\end{equation}
for $lt$ waves, which are almost electromagnetic if we assume
$k_{x}\gg k_{r}$. Using Eqs.~(\ref{Et,k}) and (\ref{Elt,k}) we can
write the diffusion coefficients for $t$ waves and $lt$ waves as
\begin{subequations}
\label{Dt}
\begin{eqnarray}
D^{ t}_{\parallel \parallel}&=&
\int_{-\infty}^{\infty }
\left(\frac{k_{\parallel}c}{\omega_{\mathbf
k}}\right)^2\frac{p_{d_\alpha}^2}{\gamma^2} I_t({\mathbf k})\,d{\mathbf k}
\label{Dtpar}\\
D^{ t}_{\perp \perp}&=&
\int_{-\infty}^{\infty }
\frac{k_{\parallel}c}{\omega_{\mathbf
k}}\frac{p_{d_\alpha}^2}{\gamma^2}
\left(1-\frac{k_{\parallel}c}{\omega_{\mathbf k}}\right)\, I_t({\mathbf
k})\,d{\mathbf k} \label{Dtper} \\
D^{ t}_{\parallel \perp}&=& D^{ t}_{\perp
\parallel}=0
\label{Dtparper}
\end{eqnarray}
\end{subequations}
for transverse $t$ waves and
\begin{subequations}
\label{Dlt}
\begin{eqnarray}
D^{ lt}_{\parallel \parallel}&=&
\int_{-\infty}^{\infty }\left(
\frac{k_{x}}{k_{\parallel }}+\frac{k_{\parallel
}c}{\omega_{\mathbf k} }\frac{p_{d_\alpha}}{\gamma
}\right)^2 I_{lt}({\mathbf k})\,d{\mathbf k}\label{Dltpar} \\
D^{lt}_{\perp \perp}&=&
\int_{-\infty}^{\infty}
\frac{p_{d_\alpha}}{\gamma }\left( \frac{k_{x}}{k_{\parallel
}}+\frac{k_{\parallel }c}{\omega_{\mathbf k}
}\frac{p_{d_\alpha}}{\gamma }\right)\nonumber \\ & &\times \left(
1-\frac{k_{\parallel  }c}{\omega_{\mathbf k} }+\frac{k_{\parallel
}c}{\omega_{\mathbf k}  }\frac{k_{x}^{2}}{k_{\parallel
}^{2}}\right)
I_{lt}({\mathbf k})\,d{\mathbf k}\label{Dltper} \\
D^{ lt}_{\parallel \perp}&=& D^{ lt}_{\perp
\parallel}=0.
\label{Dltparper}
\end{eqnarray}
\end{subequations}
for longitudinal-transverse $lt$ waves. Here we use the following
definition:
$$
I_{t,lt}({\mathbf
k})\equiv i\frac{1}{\Delta\omega_{\alpha {\mathbf k}}}\left(\frac{q_{\alpha}}{m_{\alpha}c}\right)^2\frac{E^{t,lt}_{  x}(-{\mathbf k})E^{ t,lt }_{ x}({\mathbf
k})}{V\left(
2\pi \right) ^{3}}\,. $$

Note that expressions (\ref{Dt}) and (\ref{Dlt}) are obtained
after averaging Eq.~(\ref{f0}) over the angle $\theta$. This
procedure nullifies the diffusion coefficients
$D^{t}_{\parallel\perp}$ and $D^{lt}_{\perp
\parallel}\, ;$ we also ignored $D^{lt}_{\parallel \perp}=0$
taking into account  that the terms
$$
\frac{\partial}{\partial
p_{\parallel}}\left(D_{\parallel\parallel}\frac{\partial
f_{\alpha 0}}{\partial p_{\parallel }}\right)\qquad \text{and}
\qquad \frac{1}{p_{\perp }}\frac{\partial }{\partial p_{\perp
}}\left( p_{\perp }D_{\perp \perp }\frac{\partial f_{\alpha
0}}{\partial p_{\parallel }}\right)
$$
are larger then the term
$$
\frac{\partial}{\partial p_{\parallel}}\left( p_{\perp
}D_{\parallel\perp}\frac{\partial f_{\alpha 0}}{\partial p_{\perp
}}\right)
$$
by a factor $(p_{\parallel}/p_{d_\alpha} )^2$. In Eqs.~(\ref{Dt})
and (\ref{Dlt}) $E_{x}(-\mathbf k)$ identifies the complex
conjugate to the $x$ component of electric field vector ${\mathbf
E}_{\mathbf k}$, and $V=\int^{\infty}_{-\infty}d{\mathbf r}$.

The multiplier
\begin{eqnarray}
i\frac{1}{\Delta\omega_{\alpha {\mathbf k}}}&\rightarrow&
\frac{\Gamma_{\mathbf k}}{(\Delta\omega)^2+ \Gamma_{\mathbf k}^2}
\nonumber
\\
&=&\left\{
\begin{array}{lcl}
\pi \delta (\Delta\omega), &\text{if}&
(\Delta\omega)^2 \gg \Gamma_{\mathbf k}^2; \\
1/\Gamma_{\mathbf k}, &\text{if}& (\Delta\omega)^2 \ll
\Gamma_{\mathbf k}^2.
\end{array}
\right.\label{i/Gamma}
\end{eqnarray}
presenting in the expressions of diffusion coefficients (\ref{Dt})
and (\ref{Dlt}) depends on the type of instability:  the upper case
corresponds to the kinetic approximation and the  lower case to hydrodynamic
one. Consequently they should be used  for instabilities in kinetic and
hydrodynamic approximations  respectively.

It is worth to note that the drift velocity (\ref{udr}) depends on
the Lorentz factors of the particles, $\gamma$. Hence, thermal
spread in Lorentz factors of resonant particles ($\gamma_T= |
\gamma-\gamma_0 |$) results in scatter of drift velocities,
increasing the resonant width of instability, $\Delta\omega=\omega
- k_{\parallel}v_{\parallel}-k_x u_d$. It allows to consider the
kinetic approximation of Cherenkov-drift instability. However, the
resonant width of usual Cherenkov instability is smaller than the
corresponding width of Cherenkov-drift instability. Therefore, in
the presence of narrow relativistic beam (with low value of
$\gamma_T$) the kinetic approximation for usual Cherenkov
instability is not valid. The growth rate of the instability,
$\Gamma_{\mathbf k}$, is small for hydrodynamic approximation as
well. Therefore usual Cherenkov instability, as opposed to
Cherenkov-drift instability, can not develop in relativistic
magnetized pair plasma \cite{lmmp86,elm83}.

Particle diffusion in momentum space appears both in parallel and
perpendicular directions with respect to the magnetic field $B_0$.
Diffusion process causes alteration of particle distribution
function until the quasilinear relaxation of instability is
saturated ($\partial f_0/\partial \mu t =0$, where $f_0$ is
distribution function of resonant particles). In order to
investigate the stage of saturation of quasilinear relaxation, it
is worth to rewrite Eq.~(\ref{f0}) in the following form:
\begin{equation}
\frac{\partial f_0}{\partial \mu t}=\frac{\partial}{\partial
p_{\parallel}}\left(D_{\parallel
\parallel }\frac{\partial f_0}{\partial
p_{\parallel}}\right)-\frac{1}{p_\perp}\frac{\partial}{\partial
p_{\perp}}p_{\perp}\left(D_{\perp \perp}\frac{\partial
f_0}{\partial p_{\perp}}\right).\label{D+FB}
\end{equation}
For estimation of diffusion coefficients $D_{\perp \perp}$ and
$D_{\parallel \parallel}$, we suppose that
$k_x/k_{\parallel}\approx p_d/\gamma$ and
$k_{\parallel}c\approx\omega_{\mathbf k}$ and use the following
approximations in Eqs.~(\ref{Dt},\ref{Dlt}):
\begin{eqnarray}
\left(\frac{p_{d_\alpha}}{\gamma}\right)^2 &\sim&
\left(\frac{k_{\parallel}c}{\omega_{\mathbf
k}}\right)^2\frac{p_{d_\alpha}^2}{\gamma^2}, \nonumber \\
\left(\frac{p_{d_\alpha}}{\gamma}\right)^4 &\sim&
\frac{k_{\parallel}c}{\omega_{\mathbf k}}\frac{p_{d_
\alpha}^2}{\gamma^2}\left(1-\frac{k_{\parallel}c}{\omega_{\mathbf
k}}\right), \nonumber\\ 4\left(\frac{p_{d_\alpha
}}{\gamma}\right)^2 &\sim& \left(\frac{k_x}
{k_{\parallel}}+\frac{k_{\parallel}c}{\omega_{\mathbf
k}}\frac{p_{d_\alpha }}{\gamma}\right)^2,  \nonumber \\
4\left(\frac{p_{d_\alpha}}{\gamma}\right)^4 &\sim&
\frac{p_{d_\alpha}}{\gamma}\left(\frac{k_x}
{k_{\parallel}}+\frac{k_{\parallel}c}{\omega_{\mathbf
k}}\frac{p_{d_\alpha }}{\gamma}\right) \nonumber \\
&&\times\left(1-\frac{k_{\parallel}c}{\omega_{\mathbf
k}}+\frac{k_{\parallel}c}{\omega_{\mathbf
k}}\frac{k_x^2}{k^2_{\parallel}}\right). \label{u/c}
\end{eqnarray}
Then we rewrite Eq.~(\ref{D+FB}) in the form of quasilinear
integral of Cherenkov-drift instability:
\begin{eqnarray}
\frac{\partial f_0}{\partial \mu t}&=&\frac{\partial}{\partial
p_{\parallel}}\left[\frac{A_{t,lt}}{\pi }\,\omega _{0}\gamma
_{T}^{2}\widetilde{W}^{t,lt}\left(\frac{u_d}{c}\right)^2\frac{\partial
f_0}{\partial
p_{\parallel}}\right] \nonumber \\
&+&\frac{1}{p_\perp}\frac{\partial}{\partial
p_{\perp}}p_{\perp}\left[\frac{A_{t,lt}}{\pi }\,\omega _{0}\gamma
_{T}^{2}\widetilde{W}^{t,lt}\left(\frac{u_d}{c}\right)^4\frac{\partial
f_0}{\partial p_{\perp}}\right].\label{f0t}
\end{eqnarray}
Here $\omega_0$ is the resonant frequency of $t$ and $lt$ waves
excited on Cherenkov-drift instability (see e.g. \cite{kmms91}):
\begin{equation}
\omega_0\simeq\frac{\omega _{ p}c}{\gamma _{ T}^{3/2}u_d}( \gamma
_{0}\gamma _{ p})^{1/2}; \label{wchres2}
\end{equation}
$A_{t,lt}$ is a numerical coefficient ($A_t=2$ and $A_{lt}=8$);
$\widetilde{W}^{t,lt}$ denotes the ratio of wave energy $W^{t,lt}$
and kinetic energy of plasma particles ($W_{ p}=mc^2n_{ p}\gamma_{
p}$) for $t$ and $lt$ waves respectively:
\begin{eqnarray}  \widetilde{W}^{ t,lt}=\frac{W^{t,lt}}{W_{
p}}=\int_{-\infty}^{\infty}\frac{W_{\mathbf k}^{t,lt}}{W_p}
d{\mathbf k}=\int_{-\infty}^{\infty}\widetilde{W}_{\mathbf
k}^{t,lt} d{\mathbf k}\nonumber\\
=\frac{1}{mc^2n_{ p}\gamma_{
p}}\frac{1}{V(2\pi)^3}\int_{-\infty}^{\infty}\frac{E^{
t,lt}(-{\mathbf k})E^{t,lt}({\mathbf k})}{8\pi}d{\mathbf k}
\label{Wtlt/Wp};
\end{eqnarray}
the multiplier $(i/\Delta\omega_{\alpha{\mathbf k}})$ is replaced
by $1/\Gamma_{\mathbf k}$ (\ref{i/Gamma}), where  $\Gamma_{\mathbf
k}$ is determined from Eqs.~(\ref{Gammat}) and (\ref{Gammalt})
assuming $(k_r/k_{\perp})\approx(k_x/k_{\perp})\approx 1$.

Stationary state ($\partial f_0/\partial\mu t$ = 0) is reached in
the following two cases:
\begin{description}
\item[a)] $p_d\ll p_{\perp}$. In this case the first term in
Eq.~(\ref{f0t}), containing alteration of distribution function
over parallel momenta ($\partial f_0/\partial p_{\parallel}$), is
significant. Process of quasilinear relaxation will be saturated
when plateau is formed on parallel distribution function of
resonant particles ($\partial f_0/\partial p_{\parallel}=0$);
\item[b)] $p_d \gg p_{\perp}$. In this case the second term of Eq.~(\ref{f0t}),
containing alteration of distribution function over perpendicular
momenta ($\partial f_0/\partial p_{\perp}$), appears significant.
During the process of quasilinear relaxation, energy transfers
from parallel to perpendicular motion of the particles, hence
increasing $p_{\perp}$. Relaxation will be saturated when
$p_{\perp}\sim p_d$. In this case the right hand side terms of
Eq.~(\ref{f0t}) cancel each other. Indeed, they are of the same
order but with opposite signs (for the beam particle distribution
function: $\partial f_0/\partial p_{\perp} < 0$).
\end{description}

\section{Conclusion}

In this paper we discuss the development of Cherenkov-drift
instability in relativistic magnetized pair plasma. We are taking
into account particle drift motion across the plane of SCMF lines,
which is significant for the particles of relativistic beam
penetrating the bulk pair plasma. We studied quasilinear stage of
the instability -- quasilinear interaction of excited waves over
plasma  particles and corresponding redistribution of particle
momenta. As a result, diffusion of particle momenta takes place
along, as well as across, the magnetic field lines.

The linear stage of Cherenkov-drift instability develops similarly
to the usual Cherenkov instability. The mechanism of wave
excitation is based on the well known Cherenkov wave-particle
interaction. Same as in the case of Cherenkov instability,
presence of high energy particle beam (with positive slope on the
shape of distribution function) is necessary condition for
developing of Cherenkov-drift instability. However, in the case of
Cherenkov-drift instability, the generation of oscillations with
${\mathbf E} \perp {\mathbf B}_0$ is possible only due to particle
drift motion across ${\mathbf B}_0$ (contrary to the case of usual
Cherenkov instability generating only longitudinal oscillations
with ${\mathbf E}\parallel {\mathbf k} \parallel {\mathbf B}_0$).
Cherenkov-drift instability generates both $t$ (with $E_{\perp}$)
and $lt$ (with $E_{\perp}$ and $E_{\parallel}$) waves;
electromagnetic oscillations with $E_{\parallel}$ are generated by
particle longitudinal motion with velocities $v_{\parallel}$, as
in the case of usual Cherenkov instability.

Back reaction of excited waves over resonant particles should
suppress the reason of wave excitation: in the case of
Cherenkov-drift instability, the process causes, at the same time,
formation of plateau on the distribution function of parallel
momenta (similar to the quasilinear case of usual Cherenkov
instability) and energy transfer from parallel motion of particles
to their motion across the magnetic field.

The later process, inhibiting anisotropy in momentum space, is
similar to the quasilinear relaxation of cyclotron instability.
The reason for development of cyclotron instability -- anisotropy
in momentum space ($p_{\perp}\ll p_{\parallel}$) -- is suppressed
by particle diffusion over perpendicular momenta. Perpendicular
diffusion is described by nonzero diffusion coefficients
$D_{\perp\perp}$ and $D_{\perp\parallel}$. As a result, the energy
of parallel motion of beam particles is transferring into the
perpendicular energy until $p_{\perp}\sim p_{\parallel}$. As for
Cherenkov-drift instability, relaxation is saturated since
$p_{\perp}\sim p_{d}$ and the rate of alteration of distribution
function over perpendicular momenta is described by nonzero
perpendicular diffusion coefficients $D_{\perp\perp}$
(\ref{Dtper}) and (\ref{Dltper}).

It is worth to note that, in the Cherenkov-drift instability, the
plateau on the parallel distribution function forms not only due
to parallel diffusion of particles (which transfers energy from
high velocity resonant particles to those with low velocities as
for usual Cherenkov instability), but also because of
perpendicular diffusion which transfers energy from parallel to
perpendicular motion of particles. This scenario works if the
other factors which can balance quasilinear diffusion, are not
taken into account. Such factors could be, on the one hand, the
radiation reaction force (acting on synchrotron emitting particle,
spiraling in strong magnetic field) and, on the other hand, the
force arising due to particle motion in weekly inhomogeneous
field. We plan to include these factors into consideration in the
future works.

\begin{acknowledgments}
GMa thanks D. Melrose for stimulating discussions. DSh and GMa
acknowledge hospitality of Institute of Astronomy (Zielona G\'ora
University, Poland). GMe and DSh were supported by KBN grants 2
P03D 008 19 and 5 P03D 010 21. The work was partially done at
Abdus Salam International Center for Theoretical Physics (Trieste,
Italy).
\end{acknowledgments}

\end{document}